\documentstyle[12pt]{article}

\begin{document} 
\newcommand{\be}{\begin{equation}}
\newcommand{\ee}{\end{equation}}
\newcommand{\al}{\alpha}
\newcommand{\bt}{\beta}

\title{
Practical techniques of QCD \\
vs\\ 
exact results of solvable models
}
\author{
  A.A.Penin and  A.A.Pivovarov\\
  {\small {\em Institute for Nuclear Research of the
  Russian Academy of Sciences,}}\\
  {\small {\em 60th October Anniversary
  Pr., 7a, Moscow 117312, Russia.}}
  }
\date{}
\maketitle
\begin{abstract}
We check quantitatively the validity of some popular 
phenomenological approaches of QCD in
simple models. Dispersion sum rules are considered within the ladder
approximation of a field-theoretic model with OPE given by ordinary
loop diagrams which are computable for any number of loops.
A correlator of two currents within the model complies with all
requirements of standard QCD sum rules approach for fitting 
low-lying resonances such as fast convergence and good stability  
while the exact spectrum contains no resonance. Optimized PT as it is
inspired by the principle of minimal sensitivity is analyzed within a
quantum mechanical model and
is shown to work well
as compared to pure asymptotic expansion in the coupling constant 
or Pad\'e approximation.
Renormalon technique is  
tested within another quantum mechanical
model and is found to fail to detect its low-energy structure.
\end{abstract}

\vskip 1cm

\hspace*{5mm} PACS numbers: 11.15.Bt, 11.50.Li, 11.10.Jj, 
12.38.Cy

\newpage

\section{Introduction}
With considerable and steady improvement of experimental data 
(as a recent review see, {\it e.g.} \cite{exprev})
requirements to 
the accuracy of theoretical
predictions are becoming stricter and one encounters an urgent 
necessity of quantitative
check of phenomenological methods used to treat the data 
({\it e.g.} \cite{copen}).
In the electroweak sector of the Standard Model (SM)  
perturbation theory (PT)
works well
because of smallness of the coupling constant and there is no
immediate need in its revision that was confirmed by the accurate estimate
of the top quark mass based on 
electroweak radiative corrections \cite{top}. 
For processes involving strong interactions PT calculations are now of
the comparable precision with data that caused efforts to modify
ordinary PT \cite{copen}. 
Also PT is known not to be complete due to existence of
instantons and confinement.
Several widely used approaches were suggested in QCD for describing
the properties of hadrons and were successfully used for
phenomenological predictions while their real precision has never been
checked as well as their real connection with fundamental lagrangian and
properties of QCD.
At present it seems that this latter problem will be solved only within
the lattice approximation that is going to attain a status of
ultimate judge of validity of computational methods.
Not everything is yet computable within lattice so some
phenomenological methods have no quantitative tests.
In these circumstances it is instructive to analyze some models.  

In this paper we give a quantitative analysis of several methods
used in QCD within exactly solvable
models. They are dispersion sum rules, optimized PT, and partial
resummation of PT for pinning down the non-PT effects (renormalons).
The common feature of the above techniques is an  attempt 
to obtain the non-PT information on the strong interactions 
studying asymptotic series in strong coupling constant (within ordinary PT) or 
large momentum asymptotic expansions (within operator product expansion
(OPE)). 

First we consider a two-point correlator in a field-theoretic model 
and restrict ourselves to a kind of massless $\phi^3$
ladder approximation. The spectral
density of the correlator within the approximation 
is known explicitly and does
not contain any resonances.  Meanwhile  making use of the
standard sum rules technique with a simple "resonance +
continuum" model of the spectrum predicts parameters
of the "resonance" very accurately in a sense that all
necessary criteria of stability are perfectly satisfied.
Though 
the use of sum rules implies that hadron
properties are mainly determined by several leading terms
of asymptotic expansion of the correlator of relevant
interpolating currents in deep Euclidean domain that cannot
be guaranteed by itself. This problem was studied within
several model of quantum mechanics or field theories in space-time
dimension less than four and sum rules proved to be successful
though it was stressed that there is no criterion for establishing the
validity of the technique.
In two-dimensional electrodynamics
(the Schwinger model) \cite{Schwsr}
sum rules calculations were directly checked and 
stability of the result with respect to
inclusion of higher order corrections was proposed as an
intrinsic criterion of sum rules applicability. In the
present paper we demonstrate within another exactly soluble
model an opposite example when truncation of the asymptotic
expansion of a correlator leads to missing some main
properties of its spectral density while the formal
stability requirement for corresponding sum rules is
completely satisfied. This situation reflects some general
property of sum rules and can be realized in QCD as well.
This is also the first four-dimensional field-theoretical example 
with known exact answer.

Second,  within a quantum mechanical model we study the problem of
resummation of an asymptotic perturbation series via optimization of the
perturbative expansion. In most
physically interesting models of quantum field theory
the conventional perturbation theory forms an asymptotic
series in the coupling constant which can be used for
calculation of the Green's functions only if the effective
parameter of the expansion is small enough
while for practical purposes there exist powerful methods of refining
the expansion. We also note that there are two different approaches to
such refinement. One is more mathematical and consists in dealing
directly with an asymptotic series applying methods of resummation
like Borel (with hypotheses of higher order behavior) or Pad\'e
technique.
Alternative approach relies on changing the splitting of the whole
interaction into exact part and perturbation. It is closer to Ritz
variational technique. We think that such an approach is more
physically relevant.

In the last part of our paper 
we discuss nonperturbative (power-like) corrections 
that for cases that have no simple formulation in terms
of OPE are now
mostly based on using renormalons 
(for a concise up-to-date review see \cite{ZakhAkh}).
Because the expansion parameter -- 
a running coupling constant $\al_s$ -- is sufficiently large 
for moderate energies, 
predictions differ strongly depending on a way one chooses to handle
a strong coupling constant in infrared region. 
We mimic infrared renormalons of QCD by resumming a Born series 
for $s$-wave scattering within an exactly solvable model of
quantum mechanics.
Numerical estimates show that traditional
technique ({\it e.g.} \cite{tech})
based on introduction of nonperturbative power corrections fails 
to determine the low-energy mass scale of the model
analogous to a typical resonance mass in QCD.
An alternative approach exploiting a modified
running coupling constant of the model and 
nonperturbative continuation of evolution equations into an infrared 
region gives solid and accurate 
estimate of this scale.

\section{Dispersion sum rules}
The problem of checking this technique is not new
and fairly well understood.
It is known that the solution to the sum rules problems is not
unique. 
Several toy models have been considered 
where dispersion 
sum rules 
work well. We here consider an example \cite{srpp}
focusing mainly on
the statement that the formal result is 
nice and satisfies all requirements 
of the technique but the exact spectral density is different and 
there is no intrinsic
criterion to tell us what is going wrong.
This is also a almost realistic model -- four dimensional
field-theoretical model within commonly used approximation that
reveals no explicitly pathological behavior.

\subsection{The model}
We consider a field-theoretical model in four dimensions in the 
approximation that is equivalent to ladder massless $\phi^3$ one.
In the $\phi^3$ model the expansion
parameter is dimensionless ratio of the dimensionful
coupling constant and the energy of the process in question,
so the interaction vanishes at large energies. On the other hand at low energies
the parameter of the expansion becomes large and the model
requires a nonperturbative treatment.
Thus it would be instructive to
investigate the model along the line of ordinary QCD
methods. Not to be taken quite seriously it gives
nevertheless a way to go beyond the perturbation theory
because  here  the explicit expressions for diagrams in any
order of loop expansions are known \cite{UssDav,BelUss,Brod}. The model
is attractive also because  other attempts to break the
bound of perturbation theory tend to be in dimensions
different from 4.

We modify
the usual $\phi^3$ model slightly  to make it more
convenient for our purpose.  The Lagrangian of our model
reads
\be
{\cal L}={\cal L}_0+e_1\varphi^2 A +e_2 \phi^2 A + \ldots
\label{sr1}
\ee
where ${\cal L}_0$ is a free kinetic term for all fields,
ellipsis stands for other interactions that are considered to be small
and neglected,
and we  choose $e_2=-e_1=e$.
All questions of stability
of the model (the existence of a stable ground state,
for example) remain beyond the scope of
our toy consideration.

We study a correlator of two composite
operators $j=\varphi\phi$ in the ladder approximation.
The correlator has the form \cite{BelUss}
\be
\Pi(q)=i\int\langle 0|{\rm T}j(x)j(0)|0\rangle e^{iqx}{\rm d}x,
\quad \Pi(Q^2)={1\over 16\pi^2}\ln\left({\mu^2\over Q^2}
\right)+\Delta \Pi(Q^2),\quad Q^2=-q^2,
\label{sr2}
\ee
where
\be
\Delta \Pi(Q^2) ={1\over 16\pi^2}
\sum_{L=2}^{\infty}\left(-{e^2\over 16\pi^2Q^2}\right)^{L-1}
\left(\begin{array}{c} 2L\\ L\end{array} \right)\zeta(2L-1),\quad
\left(\begin{array}{c} 2L\\ L\end{array} \right)={(2L)!\over L! L!},
\label{sr3}
\ee
where $\zeta(z)$ is the Riemann's $\zeta$-function.
Since the coupling constant
$e$ is dimensionful the expansion~(\ref{sr3})
simulates power corrections or OPE
of the ordinary QCD. 
Note that this particular subset of diagrams does not lead to infrared
problems within PT as in a 		general superrenormalizable theory.
Setting $e^2/16\pi^2=1$ we get
\be
p(Q^2)\equiv 16\pi^2\Pi(Q^2)=\ln\left({\mu^2\over Q^2}\right)+\Delta p(Q^2),
\label{sr007}
\ee
with 
\be
\Delta p(Q^2) =\sum_{L=2}^{\infty}
\left(-{1\over Q^2}\right)^{L-1} 
\left(\begin{array}{c} 
2L\\ L \end{array} \right)\zeta(2L-1)
\label{sr4}
\ee
in analogy with QCD where the scale is given and all
condensates are expressed through $\Lambda_{\rm QCD}$.

What we see first is the alternating character of the
series~(\ref{sr4}) in Euclidean direction ($Q^2>0$) due
to the special choice of interaction~(\ref{sr1}). This was
actually the reason to
create all decorations for the simple $\phi^3$ model in
which every term has the same sign for Euclidean $q$.
Note that the series (\ref{sr4}) is convergent  for $Q^2>4$ 
as it is seen from 
the asymptotic behavior of the coefficients  at large $L$
\be
\left(\begin{array}{c} 2L\\ L\end{array} \right)
\sim {4^L \over \sqrt{\pi L}}.
\label{as}
\ee

The sum~(\ref{sr4}) has the following closed form \cite{Brod}
\[
-\sum_{n=1}^{\infty}nQ^2\left[\left(1+{4\over n^2Q^2}\right)^{-{1\over
2}}-1+{2\over n^2Q^2}\right]
\]
and contains no resonances. 

The spectral density for the function
$\Delta p(Q^2)$
reads
\be
\Delta \rho(s)=-\sum_{n=1}^{\infty}
{ns\sqrt{s}\over\sqrt{{4\over n^2}-s}}\ \theta
\left({4\over n^2}-s\right)\theta (s).
\label{sr5}
\ee
Unfortunately the whole spectral density of
the correlator $\rho(s)=1+\Delta \rho(s)$ is negative
in some domains.  In case $e_1e_2>0$ it would have a
correct positive sign but the cut would be situated in the
wrong place of the complex plane (negative semiaxis).  So,
any choice of the interaction sign leads to unphysical
spectral density and the set of ladder diagrams is hardly
representative for the exact correlator 
$\langle 0|{\rm T}j(x)j(0)|0\rangle$.
The same situation is realized in QCD.
If one takes seriously the leading order correlator for
vector currents, for example, one finds that the
spectral density contains an unphysical pole at
$Q^2=\Lambda_{\rm QCD}^2$ due to running coupling constant
and does not satisfy the
spectrality condition being negative at
$s<\Lambda_{\rm QCD}^2$.
We omit these delicate points and proceed as in QCD.
Namely, whether the expansion~(\ref{sr4}) can be described
successfully with the simple formula
\be
\rho^{test}(s)=F\delta(s-m^2)+\theta(s-s_0)
\label{sr6}
\ee
for the spectral density $\rho(s)$.

An explicit expansion for the correlator reads
\be
\Delta p(Q^2)=-{6 \zeta(3)\over Q^2}
+{20 \zeta(5)\over Q^4}-{70 \zeta(7)\over Q^6}
+{252 \zeta(9)\over Q^8}
+\ldots
\label{sr7}
\ee
while the "test" form of the correlator is
\be
p^{test}(Q^2)=\ln\left({\mu^2\over Q^2+s_0}\right)+{F\over
Q^2+m^2}.
\label{sr8}
\ee

We connect expressions~(\ref{sr007}) and~(\ref{sr8}) by means of
sum rules. There are many essentially equivalent ways to extract
physical parameters from OPE (\ref{sr007}) and representation
(\ref{sr8}) however the most widely used are local duality or  
finite energy sum rules (FESR) 
\cite{FESR} and Borel technique. We check them in sequence. 

\subsection{Finite energy and Borel sum rules}
First we use FESR 
\be
\int_{0}^{s_0}s^k\rho(s){\rm d}s=
\int_{0}^{s_0}s^k\rho^{test}(s){\rm d}s
\label{sr9}
\ee
where $k=0,1,2$ because the test spectral density has
three parameters to be determined. Eqs.~(\ref{sr9}) now become
\be
F=s_0-6 \zeta(3), ~~Fm^2={1\over 2}s_0^2-20
\zeta(5), ~~Fm^4={1\over 3}s_0^3-70 \zeta(7)
%\label{sr10}
\ee
that leads to the equation for determination of
the duality threshold  $s_0$
\[
{1\over 12}s_0^4-2 \zeta(3) s_0^3+20 \zeta(5)
s_0^2 -70 \zeta(7) s_0 +420 \zeta(7) \zeta(3)-400
\zeta(5)^2=0.
\]
It has solutions $s_0=16.9$ and ${s'_0}=2.24$,
and for corresponding parameters $F=9.67$ and $m^2=12.6$
whilst $F'=-4.97$ and $m'^2=3.67$.
We will study the first solution because the second one
gives an unnatural relation between the "resonance mass" and
the "duality interval" ${s'_0} < m'^2$.

Now we check the Borel
sum  rules approach  \cite{SVZsr}. The Borel
transformation $p(M^2)$ of the function $p(Q^2)$ is
\be
p(M^2)=1-{6 \zeta(3)\over M^2}
+{20 \zeta(5)\over M^4}-{35 \zeta(7)\over M^6}+{42 \zeta(9)\over M^8}
+\ldots,
\label{sr12}
\ee
the continuum contribution gives
\be
c(M^2)={\rm exp}\left(-{s_0\over M^2}\right),
\label{sr13}
\ee
and the resonance contribution reads
\be
r(M^2)={F\over M^2}{\rm exp}\left(-{m^2\over M^2}\right).
\label{sr14}
\ee
We have plotted these functions in Fig.~1. As we see our
"test" representation  accurately simulates the asymptotic
form of the correlator at $M^2>8$.  Equating the functions
$p(M^2)$ and $r(M^2)+c(M^2)$ we obtain the Borel sum rules
to determine parameters of the resonance. The sum
rules look like ordinary QCD sum rules. Fig.~2 shows the
dependence of mass $m^2$ on the Borel variable $M^2$ with
the parameters $s_o$ and $F$ given by FESR for the
different numbers of power corrections included. For other
close values of the parameters $s_o$ and $F$ the results
are not very different from fig.~2. A stable region is
reached for $8<M^2<18$ where, on the one hand, higher order
power corrections are small and, on the other hand,
the continuum contribution 
eq.~(\ref{sr13}) does not prevail over the
resonance one eq.~(\ref{sr14}). The best stability is obtained at the
optimal values $s_o=16.6$, $F=9.39$ that shows a
consistency of the Borel sum rules and FESR. Furthermore,
the curves in fig.~2 show that inclusion of
higher order power corrections does not destroy the Borel sum
rules and even enlarges the region of stability.

Let us emphasize that the sum rules are perfectly saturated
by the artificially introduced resonance and show very good
stability though the exact spectral density does not
contained any  resonance singularities. The reason of this
phenomenon is quite transparent: using sum rules we neglect
the high order terms of large momentum expansion which do
not affect the rough integral characteristics of the
spectral density but are essentially responsible for its
local behavior. The extreme sensitivity of the local form of
the spectral density  to the high orders of perturbative
expansion  can be easily demonstrated within considered
model. For example, substituting $\zeta$-functions for $L>2$
by units in the series~(\ref{sr4}) one neglects the terms with $n>2$
in eq.~(\ref{sr5}) and gets the modified spectral density
\[
\Delta \tilde\rho(s)=-s\sqrt{s}\ \theta (s)
\left({\theta(4-s)\over\sqrt{4-s}}
+{\theta(1-s)\over\sqrt{1-s}}\right).
\]
As we see the tiny variation of the coefficients of the
asymptotic expansion leads to a drastic change of the
spectral density at low scale $s<4/9$.  Indeed, the
modified spectral density has two singular points at
$s=4~{\rm and}~1$ while the original expression has an
infinite number of singularities at  $s_n=4/n^2,~
n=1,2,\ldots$ At the same time  this variation does not
really affect the sum rules result that changes slightly
(less than $10\%$):  $\tilde s_0=17.9$, $\tilde
F=10.7$, $\tilde m^2=13.0$.

The lesson we draw is that
our toy model containing no resonances can be well fitted by
a standard "resonance + continuum" ansatz with good stability
properties. While stability criteria look appropriate 
for determining parameters of resonance in established channels
there is no intrinsic criterion of the reliability of
QCD sum rules predictions for channels where there is no experimental
data about the spectral density as, for instance, for gluonia. 
This makes sum rules predictions questionable.

Our model has recently been reanalyzed in~\cite{Ste} where some
mathematical criteria for validity of the  "resonance + continuum" ansatz
have been used in addition to a simple stability requirement. It was shown
that making use of Holder inequalities that are valid due to the positiveness
of the spectral density puts some additional constraints on
the structure of the series in $1/Q^2$ and can help in ruling out
some cases of spurious  resonances. Note however
that such a technique is not applicable in cases of
non-diagonal correlators where the spectral density
must not be positively defined.

\section{Optimized perturbation theory}
Some methods have been suggested to improve convergence
of conventional perturbation theory.
We would divide them into two groups depending on 
either one knows coefficients of expansion for $n$-term that is rare
or one knows the structure of interaction 
and makes educated guess about the best zero
order approximation.
First is a pure mathematical problem 
(like a direct summation with Borel-like technique)
and we don't touch it concentrating on the second.

A broad
class of methods is represented by general approach of
$\delta$-expansion. The basic idea of
the optimized $\delta$-expansion  is to introduce
the artificial parameter $\delta$ which interpolates between
the original theory  with Hamiltonian $H$, and
another theory, with Hamiltonian $H_0(\lambda)$ ($\lambda$
is a set of auxiliary parameters not present in the original
theory), which is soluble and reflects some main properties
of the theory we are interested in.  One defines a
new Hamiltonian depending on $\delta$
\begin{equation}
H_\delta =H_0(\lambda )+\delta (H-H_0(\lambda ))
\label{opt1}
\end{equation}
and any desired quantity is evaluated as a perturbation
series in $\delta$, which is set equal to unity at the end of
the calculations. So this parameter $\delta$ simply marks relevant
terms of the expansion developed. 
Convergence of the series is achieved
by an optimization procedure \cite{pms} {\it e.g.} by fixing
the parameters $\lambda$ at every finite order of the
expansion in $\delta$ according to principle of minimal sensitivity
(PMS) at the point where the result is least sensitive to
their variation, or the principle of fastest apparent
convergence (FAC) at the point where the next term in the
series vanishes, or some other criterion. Though the above procedure
is not rigorous it gives good numerical results in most cases. The
method has been mostly advanced in studying the anharmonic oscillator
\cite{anharm} where the convergence has been rigorously established
\cite{BDJ}. 

It seems instructive
to consider the simplest model which, nevertheless, retains most
relevant features of the real problem and, we hope, can help to
gain some intuition to cure some difficulties of PT expansion in QCD.

\subsection{The model}
The problem we will study is the ground state in the spectrum of the
stationary Schr\"{o}\-dinger equation in three dimensions \cite{pospp}
\begin{equation}
H(\alpha)\psi({\bf r})\equiv
\left(-\Delta + U(\alpha,r)\right)\psi({\bf r}) =
E\psi({\bf r}),\quad r=|{\bf r}|,
\label{opt2}
\end{equation}
where we set $2m=\hbar=1$ and
\begin{equation}
U(\alpha,r)=
\left\{
\begin{array}{ll}
-{\pi^2\over 4}-\al+\alpha^2r, &r<1,\\
 0,&r>1,
\end{array}\right.
\label{opt3}
\end{equation}
The choice of the potential is transparent:
at $\al=0$ the depth of the well is equal to the value at which the
first s-wave bound state appears.
The exact solution for the ground state energy
$E(\alpha)$  is given by the transcendental equation
\[
\sqrt{-E}=\alpha^{2/3}
{{\rm Bi}(\xi_0){\rm Ai}'(\xi_1)-{\rm
Ai}(\xi_0){\rm Bi}'(\xi_1)\over
{\rm Bi}(\xi_0){\rm Ai}(\xi_1)-{\rm
Ai}(\xi_0){\rm Bi}(\xi_1)},
\]
\begin{equation}
\xi_0=-\alpha^{-4/3}\left({\pi^2\over 4}+\alpha+E\right),\quad
\xi_1=-\alpha^{-4/3}\left({\pi^2\over 4}+\alpha-\alpha^2+E\right),
\label{opt4}
\end{equation}
where Ai(z) and Bi(z) are Airy functions \cite{mh}.
Using the asymptotic expansion
of Airy functions at large negative $\xi_{0,1}$
(small $\alpha$) we
obtain an asymptotic series for the ground state energy
\[
E(\alpha)\sim \tilde E(\alpha)\equiv -{\alpha^2\over 4}
\left\{1-\left({3\over
2}+{2\over\pi^2}\right)\alpha+\left({21\over\ 16}-{11\over
6\pi^2}+{13\over\pi^4}\right)\alpha^2-\right.
\]
\begin{equation}
\left. -\left({39\over 32}-{41\over 8\pi^2}+{35\over
6\pi^4}+{26\over\pi^6}\right)\alpha^3+\ldots\right\}.
\label{opt5}
\end{equation}
Numerically it reads
\begin{equation}
\tilde E(\alpha)= -{\alpha^2\over 4}
\left(1-1.7026\alpha+ 1.2602\alpha^2-
0.7864\alpha^3+\ldots\right),
%\label{num}
\end{equation}
and we find that the series converges fairly badly
near the point $\alpha\sim 1$. In fact, the series diverges for any
positive $\alpha$ because the coefficients of the
expansion~(\ref{opt5}) in high orders grow factorially. That reflects
the presence of a singularity of the function $E(\alpha)$ at the
origin in the complex $\alpha$ plane. The form of the singularity can
be found directly from eq.~(\ref{opt4}):  $E(\alpha)$ has a cut along
the negative semiaxis and a branch point 
at $\alpha =0$. For
sufficiently small $|\alpha|$ it is an analytical function for
$|{\rm arg}(\alpha)|<\pi$; therefore the series~(\ref{opt5}) is Borel
recoverable \cite{MarPag}, {\it i.e.} we can restore complete information
on the function $E(\alpha)$ from its asymptotic expansion. The
presence of the singularity in the Green's function of eq.~(\ref{opt2})
reflects the fact that at $\alpha=0$ the spectrum of eq.~(\ref{opt2})
changes qualitatively and a discrete part of the spectrum appears.

Consider now the model~(\ref{opt2}) within the perturbation
theory framework. At $r<1$ the potential~(\ref{opt3}) consists of two
parts: one is a constant and the other depends linearly  on $r$.
The Schr\"{o}dinger equation with the constant part of the potential
only has bound states for any positive $\alpha$ with the ground state
energy $E_0(\alpha)$ determined by the equation
\begin{equation}
\sqrt{-E_0}=\sqrt{{\pi^2\over 4}+\alpha +E_0}
\ \cot\!\left(\sqrt{{\pi^2\over 4}+\alpha+E_0}\right).
\label{opt6}
\end{equation}
If $\alpha$ is
small enough one can  consider
the constant part of the potential to be responsible for creation of
the bound state while the linear term is a perturbation because it is
suppressed by an extra power of $\alpha$. Then one can
search for the ground state energy of eq.~(\ref{opt2})
as a series in  $\alpha$ using perturbation theory.
The ground state energy of the unperturbed Hamiltonian
(the solution of eq.~(\ref{opt6}))  is an analytical
function at the origin of the complex $\alpha$ plane
and can be expanded in a
convergent series for any finite positive $\alpha$
$$
E_0(\alpha)=-{\alpha^2\over 4}
\left(1-\left({1\over 2}-{2\over\pi^2}\right)\alpha+\left({5\over\
16}-{11\over 6\pi^2}-{3\over\pi^4}\right)\alpha^2-\right.
$$
\begin{equation}
\left. -\left({7\over 32}-{13\over 8\pi^2}-{3\over
2\pi^4}-{6\over\pi^6}\right)\alpha^3+\ldots\right).
\label{opt7}
\end{equation}
However, after inclusion of perturbations
the ground state energy of the whole Hamiltonian
(the solution of eq.~(\ref{opt4})) gains the singularity
and the full series~(\ref{opt5}) becomes divergent.

\subsection{Optimized perturbative expansion}
Our purpose now is to develop the optimized perturbation theory (OPT)
for our toy model. Following the general idea of $\delta$-expansion
we have to choose the ``unperturbed'' Hamiltonian. The Hamiltonian
\begin{equation}
H_0(\alpha ')= -\Delta + U_0(\alpha',r),
\label{opt8}
\end{equation}
where
\begin{equation}
U_0(\alpha ',r)=
\left\{
\begin{array}{ll}
-{\pi^2\over 4}-\al', &r<1,\\
 0, &r>1,
\end{array}\right.
\label{opt9}
\end{equation}
is the potential of the spherical well with adjustable depth
seems to be the simplest and most appropriate choice. Here the set of
parameters $\lambda$ in eq.~(\ref{opt1}) reduces to the single
parameter $\alpha '$ characterizing the depth of the
well and we consider $0<\alpha '<2\pi^2$ so that
the trial Hamiltonian~(\ref{opt8}) has only one bound state.
Then the perturbation reads
\begin{equation}
\delta
\left( H(\alpha )-H_0(\alpha ')\right)\equiv
\delta V(\al,\al',r)=
\left\{\begin{array}{ll}
\delta(\alpha ' -\alpha+\alpha^2r), &r<1,\\
 0,      &r>1.
\end{array}\right.
\label{opt51}
\end{equation}
Expanding in $\delta$
and setting $\delta =1$ we obtain the  $n$-th order
approximation for the ground state energy as a series
\begin{equation}
E_n(\alpha,\alpha ')=E_0(\alpha ')+
E^{(1)}(\alpha,\alpha ')+\ldots +E^{(n)}(\al,\al').
\label{opt10}
\end{equation}
The zero order approximation
$E_0(\alpha ')$ (the ground state energy of the Hamiltonian~(\ref{opt8}))
is determined by transcendental equation~(\ref{opt6}) with parameter
$\al'$ instead of $\al$.
For corrections we have
$$
E^{(1)}(\alpha ,\alpha ')=\int {\rm d}{\bf r}\,
\psi_0^*(\alpha ',{\bf r})
V(\alpha ,\alpha ',{\bf r})\psi_0(\alpha ',{\bf r})\, ,
$$
\begin{equation}
E^{(2)}(\alpha ,\alpha ')=\int_0^\infty {{\rm d}E\over E_0(\alpha ')-E}
\int {\rm d}{\bf r}\, {\big |}\psi_0^*(\alpha ',{\bf r})
V(\alpha ,\alpha ',{\bf r})\psi_E (\alpha ',E,{\bf r}){\big |}^2\, ,
\label{opt52}
\end{equation}
\[
\ldots \,,
\]
where the quantity $\psi_0(\al',{\bf r})$ is the wave function
of the ground state of the Hamiltonian (\ref{opt8}) and
$\psi_E(\al',E,{\bf r})$ is the
$s$-wave function with energy
$E$ which belongs to the  continuous part of the spectrum of the
Hamiltonian (\ref{opt8}). These wave functions can be easily obtained
but the explicit expressions are too large to be presented here.

For arbitrary $\al'$  explicit analytical expressions for the
terms of the expansion~(\ref{opt10}) are absent and we have to analyze
this expansion numerically. However, if $\alpha '$ is small enough we
can expand the right hand side
of eq.~(\ref{opt10}) in the series in $\alpha '$
\[
E_0(\al')=-{\al'^2\over 4}
\left(1-\left({1\over 2}-{2\over\pi^2}\right)\al'+\left({5\over\
16}-{11\over 6\pi^2}-{3\over\pi^4}\right)\al'^2+\ldots\right),
\]
$$
E^{(1)}(\al,\al')=
{(\al'-\al)\al' \over 2}\left(1-\left({3\over4}
-{3\over\pi^2}\right)\al'+\ldots\right)+
$$
\begin{equation}
+ {\al^2\al'\over 4}\left(1+{4\over\pi^2}-\left({3\over
4}-{2\over\pi^2}+{12\over\pi^4}\right)\alpha '+\ldots\right),
\label{opt53}
\end{equation}
\[
E^{(2)}(\alpha ,\alpha ')=-\left({(\alpha '-\alpha )\over
2}\left(1+\ldots\right)
+\alpha^2\left({1\over 4}+
{1\over\pi^2}\right)\left(1+\ldots\right)\right)^2.
\]
The $\delta$-expansion~(\ref{opt10}) generates three
types of perturbative series.

Setting $\alpha '=\alpha$ in eq.~(\ref{opt10}) without  further
expansion of $E_0(\alpha)$, $E^{(i)}(\alpha,\alpha)$
in $\alpha$ we reproduce the
standard perturbation theory  of quantum mechanics with
the  unperturbed Hamiltonian $H_0(\al)$ and the perturbation
$V(\alpha,\alpha,r)$.

Expanding  $E_0(\alpha)$,
$E^{(i)}(\alpha,\alpha)$ in $\alpha$ we derive
the asymptotic series~(\ref{opt5}).
We should note that the standard analysis of the positronium spectrum
consists exactly in this machinery, {\it i.e.} every term
of perturbative expansion  around the nonrelativistic Coulomb
solution is expanded in the fine structure constant.

In our model if $\alpha'-\alpha = O(\alpha^2)$
the effective parameter of the expansion is proportional to
\begin{equation}
{1\over E_0(\al')}\int {\rm d}{\bf r}\,\psi^*(\alpha ',{\bf r})
V(\alpha ,\alpha ',r)\psi(\alpha ',{\bf r})
\sim\alpha.
\label{opt145}
\end{equation}
So $E_n(\alpha ,\alpha ')$ after expansion in
$\alpha$ correctly reproduces the first $n$ terms of
eq.~(\ref{opt5}).  Thus if we are interested in  the asymptotic
expansion only the choice of the initial approximation does not play any
role.

If, however, we intend to go beyond the asymptotic
expansion in $\al$
we should choose the ``unperturbed'' Hamiltonian in order to
provide the best convergence of the expansion. The most natural way
to optimize the expansion~(\ref{opt10}) is to
fix the  parameter $\alpha'$ in $n$-th order according to
PMS criterion at the value $\alpha^{PMS}_n$
which is defined by the equation
\begin{equation}
{\partial E_n\over\partial\alpha '}{\bigg |}_{\alpha '= \alpha^{PMS}_n}
=0\,,
\label{opt11}
\end{equation}
or according to FAC criterion at the value $\alpha^{FAC}_n$
so that
\begin{equation}
E^{(n+1)}|_{\alpha '= \alpha^{FAC}_n}=0.
\label{opt12}
\end{equation}

To solve eqs.~(\ref{opt11},~\ref{opt12}) for arbitrary $\al$
one has to proceed numerically.
However, for small $\alpha$ the roots of eqs.~(\ref{opt11},~\ref{opt12})
can be found as a series in $\alpha $. This allow us
to observe some qualitative features of OPT.
At $n=1$ using eq.~(\ref{opt53}) we find
\begin{equation}
\alpha^{PMS}_1=\alpha \left(1-\left({1\over 2}+{2\over\pi^2}\right)
\alpha+\ldots\right)\, ,
\label{opt13}
\end{equation}
\begin{equation}
\alpha^{FAC}_1=\alpha \left(1-\left({1\over 2}+{2\over\pi^2}\right)
\alpha+\ldots\right)\,.
\label{opt14}
\end{equation}
As $n\rightarrow\infty$ we have the formal series
\begin{equation}
\alpha^{\#}_\infty\sim
\alpha (1+a^{\#}_1\alpha+a^{\#}_2\alpha^2+\ldots)\,,
\label{opt15}
\end{equation}
where $\#$ stands for PMS or FAC
and the coefficients $a^{\#}_n$ are completely determined
by eqs.~(\ref{opt11},~\ref{opt12}). Moreover, by construction
\begin{equation}
E_\infty(\alpha ,\alpha_\infty^{FAC})
= E_0(\alpha^{FAC}_\infty(\alpha))\,.
\label{opt155}
\end{equation}
Power counting arguments show that
$\alpha^{PMS}_n-\alpha^{FAC}_n=O(\alpha^{n+2})$, {\it i.e.}
$\alpha^{PMS}_\infty$ and $\alpha^{FAC}_\infty$ have identical
asymptotic expansions.
Thus in the limit  $n\rightarrow\infty$ all
corrections in the optimized expansion  vanish
in our model for both the  PMS and FAC optimization prescriptions
and we obtain
\begin{equation}
E_\infty(\alpha ,\alpha_\infty^{\#})= E_0(\alpha^{\#}_\infty
(\alpha))\, .
\label{opt19}
\end{equation}
Since the function $E_0(\alpha^{\#}_\infty)$ can be expanded in a
convergent series in $\alpha^{\#}_\infty$ (eq.~(\ref{opt7})) the
singularity of the function $E(\alpha )$ at $\alpha =0$ is
absorbed by the function $\alpha^{\#}_\infty(\alpha)$, {\it i.e.}
the series~(\ref{opt15}) must be a divergent asymptotic expansion.
However, we can search for the values $\alpha_n^{\#}$ which
satisfy eqs.~(\ref{opt11},~\ref{opt12}) at every order
numerically rather than as a series in $\alpha$.

Thus we have three kinds of perturbative expansion
for the ground state energy: the asymptotic
series, the standard perturbation theory and OPT. The asymptotic
series seems to be the most primitive tool and we expect to obtain
the best results using OPT. This assumption is completely
confirmed by numerical analysis given in the next section.

\subsection{Numerical evaluation}
The results of the numerical analysis are given in
Tables~\ref{tab1}-\ref{tab3}. In Table~\ref{tab1} we present the exact value $E(\alpha)$
and the results of the asymptotic expansion $\tilde
E_n(\alpha)$ up to the $n$-th  order ($n=0,1,2,3$). In Table~\ref{tab2} we
compare the exact value $E(\alpha)$, the result of the optimized
expansion $E_1(\alpha ,\alpha_1^{PMS})$ and the results of the
standard perturbation theory up to the $n$-th order ($n=0,1$), {\it
i.e.} the values of $E_0(\alpha )$ and $E_1(\al,\al)$ not
expanded in $\alpha $.
The $[i,j]$ Pad\'{e} approximants $E^{[i,j]}(\alpha)$ with $i+j\le 3$
are shown in Table~\ref{tab3}.

We give numerical estimates for three different values of
the parameter $\alpha$ that represent typical cases:
\begin{itemize}
\item
$\alpha =0.1$, the asymptotic expansion is applicable and
justified;
\item 
$\alpha =0.5$, the asymptotic expansion is still applicable but
one has to deal with high order corrections to achieve
a satisfactory accuracy. An improvement of the asymptotic
perturbation theory is desirable;
\item
$\alpha =1.0$, the asymptotic expansion in principle can provide
only $\sim 2\%$ accuracy after summation of $\sim 13$ terms. Then the
terms start to grow.  The asymptotic
perturbation theory must be reformulated.
\end{itemize}

As we can see, a naive attempt to improve the convergence
of the perturbative expansion by simply keeping the exact
(not expanded) value of $E_n(\alpha ,\alpha )$  which sums up
some next-to-leading corrections gives a good result but is
essentially insufficient if $\alpha $ is large enough.
This shows that optimization is important for convergence.
Though we cannot directly demonstrate that without  optimization
(for any fixed $\alpha '$) the series~(\ref{opt10}) becomes divergent, as
is the case for the  anharmonic oscillator \cite{BDJ}, the numerical
analysis clearly shows the advantage of the optimized expansion.
Indeed, in all cases the best convergence is achieved within
OPT. Even for  $\alpha =1.0$ taking only
the first order correction we reach $\sim 2\%$ accuracy.  Using the
PMS prescription we can also correctly estimate the error of the
result.  For example, for $\alpha =1.0$ the naive estimate is
$(E^{(1)}(1.0,\alpha^{PMS})/E_0(\alpha^{PMS}))^2\sim 0.02$
which coincides with the real uncertainty (see Table~\ref{tab1}).

We should note that in our numerical analysis
for all values of $\alpha$
the  asymptotic series~(\ref{opt5})
is truncated far from the critical order (which depends on $\alpha$)
where it begins to diverge.  The bad convergence of the
series reveals itself only in the fact that its terms decrease quite
slowly.  On the other hand even the third order correction is hardly
available.  So the accuracy of the perturbation theory is restricted
rather by technical reasons than by the asymptotic character of the
series. Thus the optimization is not only a resummation prescription
that is useful in high orders of the asymptotic expansion but also
gives an opportunity to improve the accuracy of perturbation theory
in low orders.  This is an important benefit, especially for
nontrivial systems where high order calculations are impossible.

A remark about the Pad\'{e} approach is in order.
As we can see, some Pad\'{e} approximants are closer to the
exact result  than the plain  asymptotic expansion.
However, it is not possible to make a choice between various
approximants until the exact result  or the general structure of the
series are known. Moreover in high orders where the asymptotic
character of the series reveals itself the Pad\'{e} theory becomes
useless.  The reason is that because of their
specific structure Pad\'{e} approximants  can correctly
reproduce only pole-like singularities
while the function $E(\alpha)$ has a branch
point at $\alpha =0$. On the other hand  OPT seems to be
the most appropriate tool to deal with such a problem because it
speeds up the convergence of the perturbation series choosing the
most  natural  initial  approximation for a specific model not by
a mathematical trick. This is an automatic summation
device that does not need input information on the form of the
singularity of the bound state energy in the coupling
constant but reproduces it via optimization.  This feature can be
observed in our toy model, where the forms of the singularity of
the ground state energy at $\al = 0$ in the auxiliary and original
theories are essentially different, but OPT reproduces the correct
singular $\al$ dependence through the optimized value
of auxiliary parameter $\al'$.  This general property of
optimized perturbation theory allows one to cope even with Borel
nonsummable series \cite{BDJ} while the class of problems where the
Pad\'e theory can be successfully applied is quite restricted.

Note that perturbation theory for Green's functions is much more 
involved and one of the main problems is uniformity of 
convergence with respect to momentum.
Next section is devoted partly to this problem.

\section{Testing renormalon technique}
In this section we check ideas of method based on
notion of infrared renormalon within the
quantum mechanical model. 
The technique we further refer to as a standard one 
presently consists in
resumming bubble chains with principal value prescription for
singularities in the Borel plane. 
Rich phenomenology can be
built on such a base \cite{phen}
though the real sensitivity of the approach to the
infrared physics is unclear 
as well as an unambiguous disentangle of 
perturbative and nonperturbative (condensate)
contributions
\cite{MartSach}. Some other approaches use mostly the
modified running of the coupling constant \cite{Grun,Shirkov},
a specific recipe for scale
setting \cite{gru,BLM,Neub} or some modification of $\bt$
function to produce evolution of running coupling constant 
at small momenta
\cite{KrasPiv}. 
Initially there is no preference between these techniques
because no exact results on the behavior of PT in large orders or in IR
domain are known. Some general properties of the quantum field theory 
to be respected (like
analyticity) give no much help to distinguish between possibilities. Yet 
in phenomenological applications, the existing methods 
give different numerical results lying on the edge of experimental errors
and the final 
selection of a working frame will eventually be based on
how well the particular technique fits experimental data. 

In this section we investigate two different
approaches within a quantum mechanical model
that reflects some general features of 
renormalons.

\subsection{The model}
We consider the problem of potential scattering with 
\be 
V(r)=V_0\delta(r-r_0)
\label{pot}
\ee 
and limit ourselves to $s$-wave amplitudes \cite{renpp}.
Such a potential can be considered as a kind of confining (not
completely) one. 
We study a value of wave function at 
the origin (a free wave function is normalized to 
1).  The exact solution  for
scattering of the plain wave with 
momentum $k$ reads
\be
\psi(k)\equiv\psi(k,r=0)
=\left(1 + \frac{V_0}{k} e^{ikr_0}\sin(kr_0)\right)^{-1}.
\label{exact}
\ee 
where we set  $2m=\hbar=1$ as in the previous section. 
To study scattering of wave packages with distributed
momentum we consider an integral of the form
\be
\Psi(\lambda)=\int_0^\infty \psi(k) W(k,\lambda){\rm d}k
\label{def}
\ee 
where 
$W(k,\lambda)$ is a normalized weight function of a package depending on 
a set of parameters $\lambda$,
$\int_0^\infty W(k,\lambda){\rm d}k=1$.
It is more convenient to deal with a function $F(\lambda)$ 
\[
\Psi(\lambda) =1 + F(\lambda) 
\]
so that $F$ vanishes if the scattering potential is switched off.
Because of oscillating factors in eq.~(\ref{exact})
integrals (\ref{def}) are not well
suitable for the PT analysis (they are ``Minkowskian'' quantities). 
For $m_{}=r_0^{-1}>|V_0|$ there are no bound states in the potential (\ref{pot})
and we can carry out the Wick rotation because
$\psi(k)$ is analytic in upper semi-plane (${\rm Im}~ k >0$)
that corresponds to the physical sheet in energy $E\sim k^2$.
In ``Euclidean'' region the exact expression for $\psi(k)$ becomes
\be 
\psi(q)=\left(1+\frac{V_0}{2q}(1-e^{-2q/m_{}})\right)^{-1}
\label{eucl}
\ee 
where $k=iq$, $q>0$.
The last formula can be obtained by PT from a Born series for the
standard Lippmann-Schwinger equation of potential
scattering since we deal with a finite range potential. 
Each term of the Born series for $\psi(q)$ contains
contributions of different kinds: a power like or ``perturbative''
(modeling logarithmic terms of a QCD series) and exponentially
suppressed or ``nonperturbative'' (modeling power corrections in QCD).
We are trying to use an analogy with QCD as close as possible 
(even through terminology) though 
one must remember a toy character of the model. 
The parameter $V_0$ 
like $\Lambda_{\rm QCD}$
determines
the scale at which the perturbation theory series becomes poorly  
convergent
while $m_{}$ like $\rho$-meson mass determines the scale
of nonperturbative effects.
Note that we do not fix the sign of our parameter $V_0$ and will study 
both attractive and repulsive interaction.

At high energies ($q\gg m_{}$) the expansion parameter 
is a running coupling constant $\alpha (q)=V_0/2q$ 
(trivial asymptotic freedom as in superrenormalizable theories)
\[
\psi(q)=\psi^{as}(\al) + \psi^{np}(m_{},V_0,q)
\]
where 
\be
\psi^{as}(\al)=\sum_{n=0}^\infty(-\al)^n 
\label{naivPT}
\ee
and $\psi^{np}(m_{},V_0,q)$ stands for 
exponentially suppressed ``nonperturbative'' terms
\be
\psi^{np}(m_{},V_0,q)={V_0\over 2q}e^{-2{q}/{m_{}}}+\ldots
\label{nonPT}
\ee
Within the present model we classify terms with respect to their behavior 
at large $q$:
power like vanishing -- PT, faster than any power -- non-PT.
With the standard renormalization group (RG)
terminology we have 
\begin{equation}				
\bt(\al)=q{\partial\al\over \partial q}=-\al.
\label{b1}
\end{equation}
Resummation in eq.~(\ref{naivPT}) 
(in the spirit of RG) results in definition of 
a new (renormalization group improved) ``running'' coupling constant
\be
\al_{as}(q) = {V_0\over 2q+V_0}
\label{resumPT}
\ee
with a $\bt$ function
\be
\bt^{as}(\al_{as})=q{\partial\al_{as}\over \partial q}=-\al_{as}(1-\al_{as}).
\label{bas}
\ee
The use of this expansion parameter allows us to improve
the perturbation theory and to sum up all ``perturbative'' 
power terms of the series (\ref{naivPT})
\be
\psi^{as}(q)=1+\al_{as}(q).
\label{resumfn}
\ee

Now we turn to consideration of the wave 
package of a specific form
given by the following weight function
\[
W(q,Q)=Q{e^{-{Q/q}}\over q^2}.
\]
This weight function has a bump of the width $\sqrt{3}Q$ 
located at $q\sim Q/2$
so the above  wave package can be considered as a ``probe''
of the scattering potential at the scale $\sim 2/Q$.

\subsection{Borel resummation and modified $\bt$ function}
It is easy to see that our observable (\ref{def})
suffers from the renormalon.
Substituting $\psi(q)$ in eq.~(\ref{def}) by its asymptotic 
expansion (\ref{resumfn}) we obtain 
\be
F^{as}(\al)= \int_0^\infty \al_{as}(q)W(q,Q){\rm d}q.
\label{df}
\ee
The quantity $F^{as}(\al)$ has a typical structure 
of QCD observable containing renormalon, {\it i.e.}
it is an integral of some weight function multiplied by
a running coupling constant over the interval that
includes strong coupling domain.
Note that in our model the use of the running coupling 
constant (\ref{resumPT})
in the integrand accounts for all perturbative corrections. One can only
dream about that in QCD where a trick based on
a pure assumption motivated by the ``naive nonabelianization''\cite{nonab} 
is used.

After integration
we get the series with factorially growing coefficients
\be
F^{as}(\al)=\sum_{n=1}^\infty (-1)^nn!\al^n.
\label{serf}
\ee
Properties of the series (\ref{serf}) depend crucially
on the sign of $\al$ or $V_0$.  Let us consider first repulsive 
potential $V_0>0$.
Then the alternating sign series~(\ref{serf}) is Borel summable
(in QCD it corresponds to an ultraviolet renormalon).
The Borel image
\be
\tilde F^{as}(u)=-{u\over 1+u} 
\label{serfb}
\ee
has a pole at $u=-1$ and is a regular function
on the positive semiaxis. 
So the Borel summation leads to an unambiguous result 
($\al=V_0/2Q$)
\be
F_B(\al)= {e^{-{1/\al}}\over\al}{\rm E_1}\left({1\over\al}\right)-1 
\label{bsp}
\ee
where ${\rm E_1(x)}$ is the integral exponent~\cite{mh}
$$
E_1(x)=\int_x^\infty{e^{-t}\over t}{\rm d}t.  
$$
One could naively hope that the Borel summation provides us
with nonperturbative information on the function $F(Q)$ and on 
the magnitude of nonperturbative parameter $m_{}$.
However numerical analysis shows that 
$F_B(Q)$ does not approximate well
the exact function $F(Q)$ for intermediate $Q\sim m_{}$ (Fig.~3). 
Indeed,  
$\al_{as}(q)$ is the best expansion parameter (exponentially accurate)
at large momenta
but $\psi^{as}(q)$ does not approximate 
well the function $\psi(q)$ at small $q$
and eq.~(\ref{bsp}) tells us nothing about the 
parameter $m_{}$ that measures exponentially suppressed 
``nonperturbative'' contributions.
Alternatively, one can compute the function $F(Q)$ within 
the modified perturbation theory with sufficient accuracy even at 
very small $Q$. For this purpose one has to choose
a relevant expansion parameter. First we write
\[
\al_{as}(q)
=
\al^{(1)}_\mu(q)\left(1-{\mu-V_0\over V_0}\al^{(1)}_\mu(q)\right)^{-1}
=
\al^{(1)}_\mu(q)
\sum_{n=0}^\infty\left({\mu-V_0\over V_0}\right)^n\al^{(1)}_\mu(q)^n,
\]
$$
\al^{(1)}_{\mu}(q) = {V_0\over 2q+\mu}
$$
where $\mu$ is a parameter.
Now we limit ourselves to only two terms of this expansion that is reasonable 
in PT region and define new expansion parameter
\be
\al_{\mu}^{(2)}(q) 
=\al^{(1)}_\mu(q)\left(1+{\mu-V_0\over V_0}\al^{(1)}_\mu(q)\right)
={V_0\over 2q+\mu}\left(1+{\mu -V_0\over 2q+\mu}\right)
\label{resumPTm}
\ee
with a $\bt$ function
\[
\bt^{(2)}_\mu(\al)=-\al +O(\al^2)
\]
that is a PT transformation.
The perturbation theory series for $\psi(q)$ in $\al_{\mu}^{(2)}$
reads 
\be
\psi^{as}(q)=1+\al_{\mu}^{(2)}(q)+O(\al(q)^2).
\label{resumm}
\ee
Taking the first order term in 
eq.~(\ref{resumm}) and fixing the parameter $\mu$ 
at some value we find the function 
$F^{\mu}(Q)=\int_0^\infty \al_{\mu}^{(2)}(q)W(q,Q){\rm d}q$ 
to be very close to the 
exact function $F(Q)$ up to the very small $Q$ (Fig.~3).
Note that at very large $Q$ the function $F_B(Q)$ becomes 
closer to exact result $F(Q)$ than $F^{\mu}(Q)$  
because  $F_B(Q)$ and $F(Q)$  
have the same asymptotic expansion. On the other hand
lifting this too strong condition
we get the function $F^{\mu}(Q)$ that approximates the exact function $F(Q)$
uniformly for all finite $Q$.
Because at the optimal $\mu$ the function $\psi^{(2)}(q)$
is close to  $\psi(q)$ for all finite $q$ this approximation is universal
in a sense that it works well for various forms of 
scattering package.    
The optimal value $\mu^{opt}(m_{},V_0)$ can be extracted from experiment
(in our model the exact solution plays the role
of experimental data). It turns out to be very sensitive to the 
variation of $m_{}$. So in this way we obtain the real information 
on the scale $m_{}$ at which nonperturbative effects become important. 
    
The case of attractive potential $V(r)=-V_0\delta(r-r_0)$, $V_0 >0$ 
at first sight seems to be completely different. 
The running coupling constant 
\[
\al_{as}(q)={V_0\over 2q-V_0}
\]
with $\bt$ function 
$
\bt(\al)=-\al(1+\al)
$
has a ``Landau pole'' at $q=V_0/2>0$.  
The series (\ref{serf}) now becomes
\be
F^{as}(\al)=\sum_{n=1}^\infty n!\al^n
\label{pos}
\ee
and it is not Borel summable: we encounter
an ``infrared renormalon''.
Its Borel image has a pole on the positive 
semiaxis in full analogy with QCD (when the QCD $\bt$ function
is taken in one loop approximation). 
So Borel procedure leads to an ill-defined representation in case of 
attractive potential.
Standard technique could consider it as a signal of the presence 
of nonperturbative contributions. One should stress however that 
exact function undergoes no qualitative change in the low energy domain
(see also \cite{deRaf}).
Following the line of QCD renormalon technique
we define the result 
of Borel summation by deforming the integration contour 
in the complex $u$ plane.
The result obtained in this way depends on the specific form of
an integration contour
while an appropriate nonperturbative
part (condensates or power corrections in QCD) must cancel 
this dependence.
In our model we use the principal value (PV) prescription 
to define the sum of the series (\ref{pos})
\be
F_B(\al)= {\rm PV}\int_0^\infty e^{-{u/\al}}{u\over 1-u}{{\rm d}u\over\al}=
{e^{-{1/\al}}\over\al}{\rm Ei}\left({1\over\al}\right)-1  
\label{bsm}
\ee
where Ei(x) is an integral exponent~\cite{mh}
\[
Ei(x)={\rm PV}\int^x_{-\infty}{e^{t}\over t}{\rm d}t,
\]
$\al=V_0/2Q>0$.  
Within the renormalon picture one should search for the exact 
function $F(Q)$ in the form  
\be
F(Q)=F_B(\al)+Ce^{-{1/\al}}+\ldots  
\label{bsmc}
\ee
Here the first term has the same asymptotic expansion as 
the exact function $F$, a constant 
$C$ gives the leading exponentially suppressed 
correction and ellipsis stands for ``higher twist'' contributions.
The power of the exponent in~(\ref{bsmc}) is 
determined by the position of the pole in the Borel image.
To find the value of $C$ one has to use purely
nonperturbative method or extract it from experiment.
Note that the 
form of first term, the value of the constant $C$, 
and high order corrections
do depend on summation prescription while the whole
sum does not by construction. 
The parameter $C(m_{},V_0)$ is quite sensitive to the 
variation of $m_{}$ so it can be considered as a ``probe''
of nonperturbative effects what is the main issue of the 
renormalon calculations.  
However, we find (Fig.~4 that
eq.~(\ref{bsmc}) poorly approximates
the exact function $F(Q)$ for $Q\sim m_{}$. 
The reason is the same as in the case of
Borel summable series. Namely, 
$\al_{as}(q)$ is a bad expansion parameter at small momenta. 
The only difference with previous case is that for Borel nonsummable series
the running coupling $\al_{as}$ becomes singular at some point
and in principle can not be used for approximation of regular
function $\psi(q)$. Stress once again however that it still accumulates
all PT terms exactly as in the previous case. 
Moreover eq.~(\ref{bsmc}) has no relation to the exact function $F(Q)$  
because the leading non-PT term in the  expansion of  $F(Q)$ at large $Q$
has completely different $\al$ dependence
\[
\sqrt{\al V_0\over m_{}}
\exp\left(-2 \sqrt{V_0\over \al m_{}}\right).
\]
As a consequence, the parameterization~(\ref{bsmc})
is not universal {\it i.e.} one  gets essentially different values of C 
for different  scattering packages.   

Again, introducing a running coupling constant of the 
form (\ref{resumPTm}) with an appropriate $\mu$
we obtain a uniform approximation of the function $F(Q)$
for all $Q$ (Fig.~4). 
In QCD this prescription corresponds to the use
of (probably mass dependent) RG equation for 
strong coupling constant with infrared regular 
solution~\cite{Grun,KrasPiv,Khoz}. 

Note that in this way as within renormalon picture we determine the
scale where nonperturbative effects become crucial 
rather than find the exact form of  $F(Q)$. However, the use
of an appropriate expansion parameter allows us 
to achieve much higher accuracy at intermediate momenta in 
calculations of the integral observables like $F(Q)$.

\section{Conclusion}
While it is clear that sum rules 
should be applied with great care to cases 
where the physical spectrum is not known we found that there is no
intrinsic criterion within the technique for detection that something
goes wrong. All criteria of convergence, stability, existence of a mass gap are
perfectly satisfied. Still this is not completely artificial
construction but a model based on an approximation of quantum field
theory. 

The optimized perturbation theory shows that knowledge of the
potential allows to reduce the necessary number of PT terms to a bare minimum
still providing good accuracy.
We would like to stress that in our model we used more than simple
play with a finite piece of PT series,
we chose the leading approximation differently. 
In QCD, however, the choice of zero order approximation 
is practically unique because there is no other solvable 
approximation but free relativistic field
theory and no real optimization is therefore possible at present.
In this respect a simple change of a renormalization prescription
has no deep physical justification though can be quite successful
phenomenologically in some particular cases. 

The analysis of two possible ways of extracting information on low
energy domain of a quantum mechanical model
shows that approach based on optimization of the PT through introduction
of a flexible expansion parameter using 
the freedom of choice of the scheme
is more efficient for moderate energies than
direct Borel resummation technique.
In our model also the singularity of Borel image for attractive
potential
does not correspond to leading non-PT asymptotics of the exact function
that is one of main reasons for using the renormalon technique in
phenomenological applications of QCD.
Though obtained in a toy model, these observations may serve as a
ground for using a modified running of the coupling constant of QCD
in the
infrared domain for phenomenological applications instead
(or in addition to) the 
standard renormalon technique.

\vspace{10mm}

\noindent
{\large \bf Acknowledgments}

\vspace{1mm}

\noindent
We thank V.A.Matveev for support and encouragement,
V.A.Rubakov for his
interest in the work and discussions.
We thank
participants of INR Theory Division seminar
for comments and useful remarks. We are grateful to
P.M.Stevenson for correspondence.  
The work of A.A.Pivovarov is supported in part
by Russian Fund for Basic Research grant N~96-01-01860 and N~97-02-17065.
The work of A.A.Penin is supported in part by INTAS grant N~93-1630-ext
and N~93-2492-ext (research program of 
International Fund for Fundamental Physics in
Moscow) and Russian Fund for Basic Research grant N~97-02-17065.

\newpage
\vspace{2cm}
\begin{center}
{\Large \bf Tables}
\end{center}
\vspace{5mm}
\begin{table}[h]
\begin{center}
\begin{tabular}{|c|c|c|c|c|c|}
\hline
$\alpha$ & $E(\alpha)$ & $\tilde E_0(\alpha)$ & 
$\tilde E_1(\alpha)$ & $\tilde E_2(\alpha)$ & $\tilde E_3(\alpha)$ \\ \hline
0.1      & 0.2101      & 0.2500                 
& 0.2074             & 0.2105               & 0.2104               \\ \hline
0.5      & 2.426       & 6.250                  
& 0.9292             & 2.898                & 2.284                \\ \hline
1.0      & 2.144       & 25.00                  
& -17.57             & 13.94                & -5.721               \\ \hline
\end{tabular}
\end{center}
\caption{ The exact ground state energy $E(\alpha)$ and
the result of  the asymptotic expansion $\tilde
E_n(\alpha)$ up to the $n$-th  order ($n=0,1,2,3$)
(in units  of $10^{-2}$).}
\label{tab1}
\end{table}

\vspace{5mm}

\begin{table}[h]
\begin{center}
\begin{tabular}{|c|c|c|c|c|}
\hline
$\alpha$ & $E(\alpha)$ & $E_1(\alpha,\alpha_1^{PMS})$ & $E_0(\alpha)$ & $E_1(\alpha,\alpha)$  \\ \hline
0.1      & 0.2101      & 0.2104                       & 0.2430        & 0.2092                \\ \hline
0.5      & 2.426       & 2.424                        & 5.531         & 1.915                 \\ \hline
1.0      & 2.144       & 2.116                        & 20.13         & -4.200                \\ \hline
\end{tabular}
\end{center}
\caption{ The exact ground state energy $E(\alpha)$, 
the result of the optimized
expansion up to the  first order $E_1(\alpha,\alpha_1^{PMS})$
and the results of the standard perturbation theory up to the $n$-th
order $E_n(\alpha,\alpha)$ $(n=0,~1)$ (in units  of $10^{-2}$).}
\label{tab2}
\end{table}

\vspace{5mm}

\begin{table}[h]
\begin{center}
\begin{tabular}{|c|c|c|c|c|c|c|}
\hline
$\alpha$ & $E(\alpha)$ & $E^{[1,1]}(\alpha)$ & $E^{[0,2]}(\alpha)$ & $E^{[2,1]}(\alpha)$ & $E^{[1,2]}(\alpha)$  & $E^{[0,3]}(\alpha)$ \\ \hline
0.1      & 0.2101      & 0.2104              & 0.1874              & 0.2104              & 0.2104               & 0.2104              \\ \hline
0.5      & 2.426       & 2.366               & 2.340               & 2.430               & 2.424                & 2.562               \\ \hline
1.0      & 2.144       & 0.5388              & 5.759               & 1.833               & 1.600                & 4.331               \\ \hline
\end{tabular}
\end{center}
\caption{The exact ground state energy $E(\alpha)$ and various $[i,j]$
Pad\'{e} approximants $E^{[i,j]}(\alpha)$ with $i+j\le 3$ (in units  of 
$10^{-2}$).} 
\label{tab3}
\end{table}

\newpage
\begin{center}
{\Large \bf Figure Captions}
\end{center}
\vspace{5mm}

\noindent
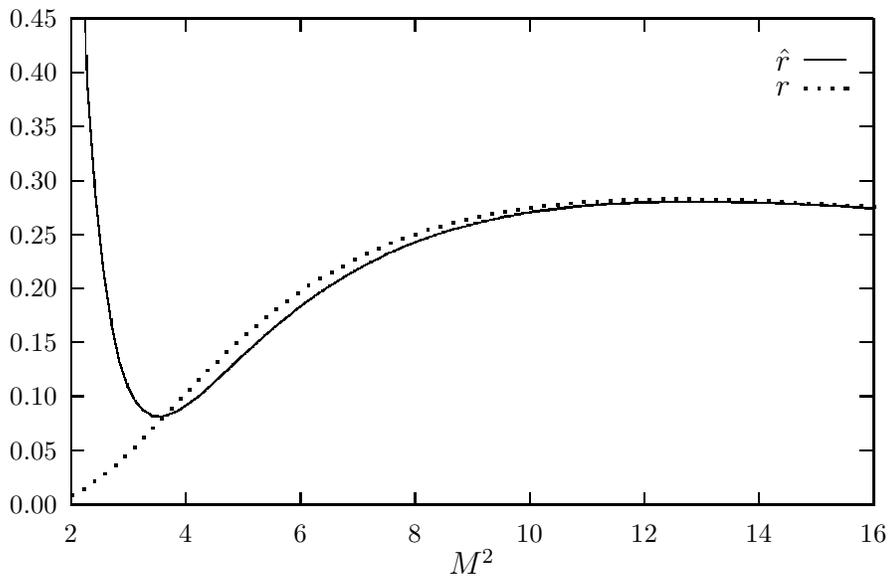
Fig.~1.  The resonance contribution $r(M^2)$ to the Borel
sum  rules  and the functioin
$\hat r(M^2)=p(M^2)-c(M^2)$.\\[2mm]
Fig.~2.  The mass $m^2$  plotted as a function of
Borel variable $M^2$ with two (a), three (b) and four (c)
orders of power corrections included. The arrows mark
stability interval of the Borel sum  rules.\\[2mm]
Fig.~3. Numerical results for repulsive potential $V_0=1$,
$m_{}=3$.\\
Function $F^{\mu}(Q)/F(Q)$, the optimal value $\mu =6$ (line $a$).\\
Function  $F_B(Q)/F(Q)$ (line $b$).\\[2mm]
Fig.~4. Numerical results for atractive potential $V_0=1$, $m_{}=3$.  \\
Function  $F^{\mu}(Q)/F(Q)$ for the optimal value $\mu =3.5$ (line $a$).\\
Function  $F_B(Q)/F(Q) $ (line $b$).\\
Function  $(F_B(Q)+Ce^{-{1/\al}})/F(Q)$ 
for an optimal value $C=-0.06$ (line $c$).

\newpage
\vspace*{50mm}
% GNUPLOT: LaTeX picture
\setlength{\unitlength}{0.240900pt}
\ifx\plotpoint\undefined\newsavebox{\plotpoint}\fi
\sbox{\plotpoint}{\rule[-0.200pt]{0.400pt}{0.400pt}}%


\vspace{5mm}
\begin{center}
{\bf Fig. 4}
\end{center}


\begin{thebibliography}{99}
\bibitem{exprev}G.Altarelli, CERN-TH/96-265, hep-ph/9611239.
\bibitem{copen}V.Braun and L.Magnea, 
Renormalons and Power Corrections in QCD, NORDITA - 96/64 P,
Proc. of the workshop,
Nordita, Copengagen, August 2-4, 1996.
\bibitem{top}F.Abe et al, (CDF) Phys.Rev.Lett. {\bf 74}(1995)2626,\\
S.Abachi et al, (D0) Phys.Rev.Lett. {\bf 74}(1995)2632.               
\bibitem{Schwsr}A.A.Pivovarov, N.N.Tavkhelidze and V.F.Tokarev,
Phys.Lett. {\bf B132}(1983)402;\\
A.A.Pivovarov, A.N.Tavkhelidze and V.F.Tokarev,
Theor.Math.Phys. {\bf 60}(1985)765;\\
A.A.Pivovarov and V.F.Tokarev, Yad.Fiz. {\bf 41}(1985)524.
\bibitem{ZakhAkh}R.Akhoury and V.I.Zakharov, hep-ph/9610492.
\bibitem{tech}P.Ball, M.Beneke and V.M.Braun, Nucl.Phys. {\bf B452}(1995)563.
\bibitem{srpp}A.A.Penin and A.A.Pivovarov, Phys.Lett. {\bf B357}(1995)427.
\bibitem{UssDav} N.I.Ussukina and A.I.Davydychev, Phys.Lett. 
{\bf B305}(1993)136.
\bibitem{BelUss}V.V.Belokurov and N.I.Ussukina, J.Phys. {\bf A16}(1983)2811.
\bibitem{Brod} D.J.Brodhurst, Phys.Lett. {\bf B307}(1993)132.
\bibitem{FESR} N.V.Krasnikov and A.A.Pivovarov,
Phys.Lett {\bf B112}(1982)397;\\
N.V.Krasnikov, A.A.Pivovarov and N.N.Tavkhelidze,
Z.Phys. {\bf C19}(1983)301.
\bibitem{SVZsr}M.A.Shifman, A.I.Vainshtein and V.I.Zakharov,
Nucl. Phys. {\bf B147}(1979)385.
\bibitem{Ste} T.G.Steele, S.Alavian, J.Kwan,  Phys.Lett. {\bf B392}(1997)189. 
\bibitem{pms}P.M.Stevenson, Phys.Rev. {\bf D23}(1981)2916, 
Nucl. Phys. {\bf B231}(1984)65.
\bibitem{anharm}W.E.Caswell, Ann.Phys.(N.Y.) {\bf 123}(1979)153;\\
J.Killingbeck, J.Phys. {\bf A14}(1981)1005.
\bibitem{BDJ}I.R.C.Buckley, A.Duncan and H.F.Jones, Phys.Rev. 
{\bf D47}(1993)2554;\\ 
A.Duncan and H.F.Jones, Phys.Rev. {\bf D47}(1993)2560.
\bibitem{pospp}A.A.Penin and A.A.Pivovarov, Phys. Lett. {\bf B367}(1996)342.
\bibitem{mh}Handbook of mathematical functions, ed. by M.Abramovitz
and I.A.Stegun, National Bureau of standards (1964).
\bibitem{MarPag}W.Marciano and H.Pagels, Phys. Rep. {\bf 36}(1978)137.
\bibitem{phen}
A.V.Manohar and M.B.Wise, Phys.Lett. {\bf B344}(1995)407;\\
B.R.Webber, Phys.Lett. {\bf B339}(1994)148;\\ 
Yu.L.Dokshitzer and B.R.Webber, Phys.Lett. {\bf B352}(1995)451;\\
I.I.Bigi, M.A.Shifman, N.G.Uraltsev, 
A.I.Vainshtein, Phys.Rev. {\bf D50}(1994)2234;\\
M.Beneke and V.M.Braun, Nucl.Phys. {\bf B454}(1995)253;\\
R.Akhoury and V.I.Zakharov, Nucl.Phys. {\bf B465}(1996)295.
\bibitem{MartSach}G.Martinelli and C.T.Sachrajda, Nucl.Phys. {\bf B478}(1996)660.
\bibitem{Grun}G.Grunberg, Phys.Lett. {\bf B372}(1996)121.
\bibitem{Shirkov}D.V.Shirkov, I.L.Solovtsov, hep-ph/9604363.
\bibitem{gru}G.Grunberg, Phys.Lett. {\bf B95}(1980)70.
\bibitem{BLM}S.J.Brodsky, P.G.Lepage and P.B.Mackenzie,   
Phys.Rev. {\bf D28}(1983)228;\\ 
S.J.Brodsky, H.J.Lu, Phys.Rev. {\bf D51}(1995)3652.
\bibitem{Neub}M.Neubert, Phys.Rev. {\bf D51}(1995)5924.
\bibitem{KrasPiv}N.V. Krasnikov and A.A. Pivovarov, hep-ph/9510207,
hep-ph/9512213, Mod.Phys.Lett. {\bf A11}(1996)835. 
\bibitem{renpp}A.A.Penin and A.A.Pivovarov,  hep-ph/9612204,
Preprint KEK 96-144.
\bibitem{nonab}D.J.Broadhurst, Z.Phys. {\bf C58}(1993)339;\\
D.J.Broadhurst and A.G.Grozin, Phys.Rev. {\bf D52}(1995)4082;\\
M.Beneke, V.Braun, Phys.Lett. {\bf B348}(1995)513;\\
C.N.Lovett-Turner, C.J.Maxwell, Nucl.Phys. {\bf B452}(1995)188.
\bibitem{deRaf}S.Peris, E. de Rafael, Phys.Lett. {\bf B387}(1996)603.
\bibitem{Khoz}Yu.L.Dokshitser, V.A.Khoze and S.I.Troyan, 
Phys.Rev. {\bf D53}(1996)89. 
\end{thebibliography}
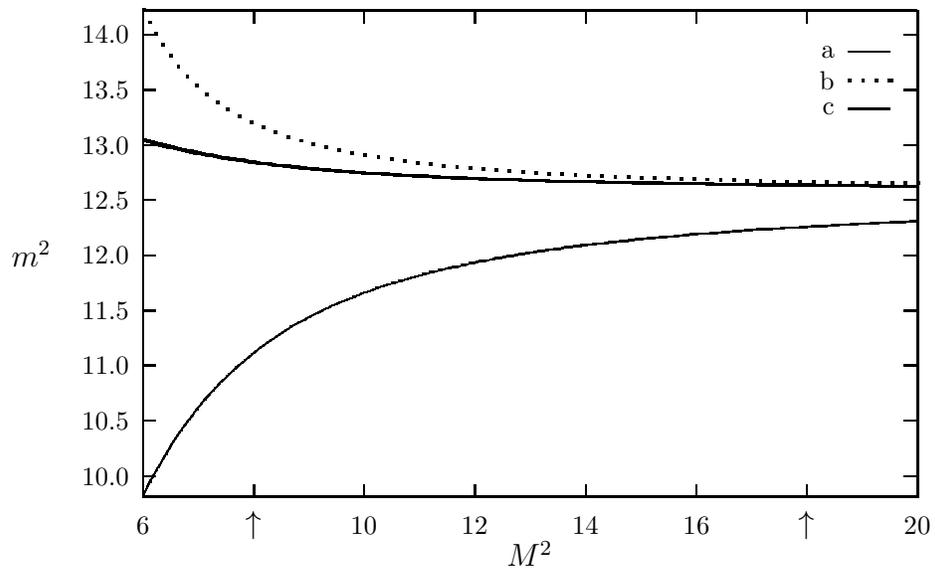
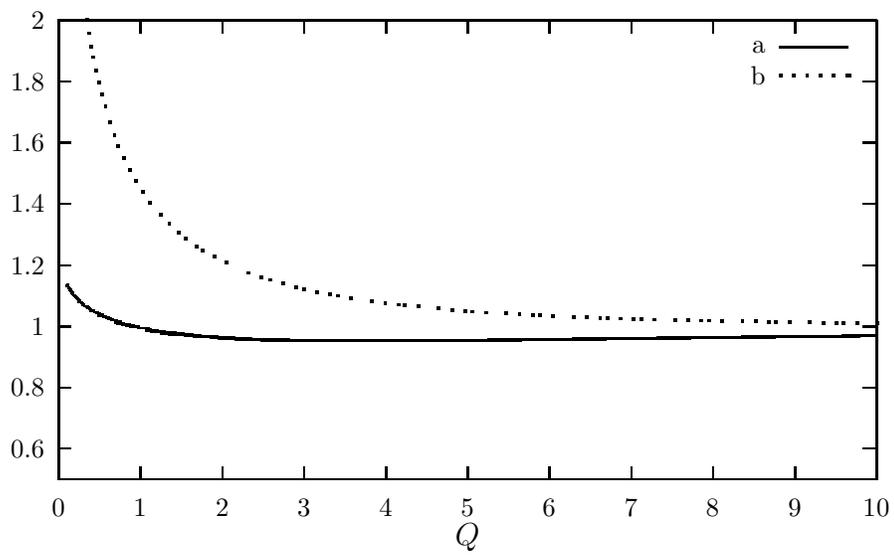
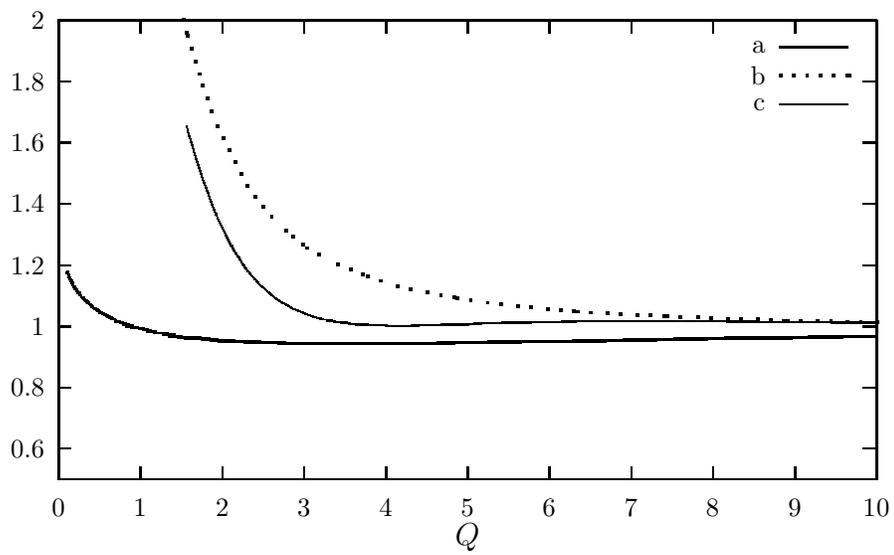
\end{document}